\documentclass[twocolumn,showpacs,preprintnumbers]{revtex4}
\usepackage{amsmath}
\usepackage{epsfig}
\usepackage{graphicx}

\begin{document}

\title{Tamm's surface states and 
%Gap effect on the 
Bose-Einstein condensation}
% transici\'on
%a la fase condensada de BE}
\author{M.A. Sol\'{\i}s}
\affiliation{Instituto de F\'{\i}sica, Apartado Postal 20-364, Universidad Nacional Aut\'onoma de M\'exico, M\'exico City, MEXICO }
\date{Modified: \today / Compiled: \today}

\begin{abstract}
We calculate and discuss the effects on the thermodynamic properties of a 3D Bose gas caused by a gap $\Delta$ in the energy of the particles constituting the gas that without the gap behaves like an ideal Bose gas. 
Explicit formulae with arbitrary $\Delta$ values are discussed
for: the critical temperature which increases as the gap grows; the condensate fraction; the internal energy; and the constant-volume specific heat found
to possess a jump-discontinuity for any $\Delta$ different from zero. Three dimensional infinite ideal Bose gas results are recovered when we the energy gap goes to zero. 

PACS: 03.75.Fi; 05.30.Jp; 67.40.Kh \newline
Keywords: Bose-Einstein condensation; phase-transition; thermodynamic
properties.

\end{abstract}

\maketitle

\section{Introduction}

After the discovery of helium superfluidity by Kapitza and Allen independently \cite{Kapitza,Allen}, London \cite{London} proposed that this novel phenomenon could be a consequence of 
%[was a consequence of] 
the Bose-Einstein condensation(BEC) predicted by Einstein in 1925 \cite{BEC}.
London himself calculated the critical temperature of a
3D ideal Bose gas with mass and density equal to that of helium four; the critical temperature
obtained is 3.2 K, which is very close to the
temperature at which helium four becomes superfluid, i.e., 2.17 K.   
This result feed the idea that helium superfluidity was motivated by a BEC of the helium atoms. However, the specific heat of helium four near absolute zero temperature had been observed and measured \cite{Keesom} to increase exponentially with temperature which is not expected for a 3D ideal Bose gas whose specific heat increases proportional to $T^{3/2}$.   To reproduce this exponential behavior with temperature, an artificial gap in the dispersion relation was introduced \cite{London2,japoneses}. Although the exponential behavior is reproduced for low temperatures, the critical temperature increases beyond the Bose-Einstein critical temperature when the energy gap increases. This behavior is contrary to what is observed in helium, where the interaction reduces the superfluidity critical temperature below the 3D BEC critical temperature when the density is that of liquid helium four. Although many theoretical efforts \cite{gapjustification} have been made to justify the gap as a result of interactions among the particles, the gap not only has not showed up but there are arguments against its existence \cite{HP}.

Now days, when it is possible to reproduce almost any space lattice structure trapping fermions or bosons in the laboratory, we consider relevant to reconsider
%go back to think in 
the surface effect on the dispersion relation of quantum particles placed in optical periodic lattices which simulate crystalline solids. Many year ago, after the publication of the Kronig-Penney model \cite{KP} for an infinite solid, Tamm \cite{Tamm} discussed the finite size effect on the particle energy spectrum using a semi-infinite idealized one-dimensional crystal. Tamm reported the occurrence of an isolated energy level for each allowed energy band. These isolated surface level are called Tamm's surface states. 

In this article we return to the London's idea to introduce a gap in the energy spectrum of bosons but keeping in mind the existence of Tamm surface states, to give details of the energy gap effects on the Bose-Einstein critical temperature, its isochoric specific heat as a function of temperature as well as the its specific heat jump at $T_c$.              

In Sec. II after definition of the system, we calculate: a) the critical temperature as a function of gap; b) condensed fraction. In Sec. III we give the following thermodynamic properties:the internal energy, the isochoric specific heat as a function of temperature and the  jump height of the specific heat at $T_c$. In Sec. IV we give our conclusions. 

 \section{Bose gas with dispersion relation plus a gap}
 
Our system is a 3D infinite Bose gas of $N$ particles of mass $m$ whose energy as a function of its momentum $hbar k$, i.e. its dispersion relation,  is given by    
\begin{equation}
\varepsilon _{k}=\left\{
\begin{array}{c}
\varepsilon _{0}{\ \ \ \ \ \ \ \ \quad \ \quad \ \quad \ \ \ \ \ \mbox{if}\ \ \ \ }k=0 \\
\varepsilon _{0}+\Delta +\hbar^{2}k^{2}/2m{\ \ \ \mbox{if} \ \ \ }k>0%
\end{array}%
\right. ,  
\label{rdg}
\end{equation}

For a finite temperature the particle number $N$ is distributed between the energy ground state and the excited states, i.e., $N=N_0(T)+ N_e(T)$ with 
\begin{equation*}
N_{0}(T)={\frac{1}{e^{\beta \lbrack \mathbf{\varepsilon }_{0}-\mu (T)]}-1},}
\label{N0}
\end{equation*}
the particles in the energy ground state and
\begin{equation}
N_e = \sum_{\mathbf{k\neq 0}}n_{k}  = \sum_{\mathbf{k\neq 0}}{\frac{1}{e^{\beta \lbrack \mathbf{\varepsilon }%
_{k}-\mu (T)]}-1}} \ ,  
\label{Ne1}
\end{equation}
the particles in the excited states,    
with $\beta \equiv 1/k_{B}T$, y $\mu (T)\leq \varepsilon _{0}$ the chemical potential.

In the thermodynamic limit sums in (\ref{Ne1}) are well approximated by integrals and after defining $\xi \equiv \beta \hbar ^{2}k^{2}/2m$ then $d \xi =(\beta \hbar^{2}/2m)2k \ dk$ and $\xi ^{1/2}=(\beta \hbar
^{2}/2m)^{1/2} \ k$, Eq. (\ref{Ne1}) becomes
\begin{eqnarray}
N_e &=&\frac{L^{3}(2m/\hbar ^{2})^{3/2}}{4\pi ^{2}\beta^{3/2}}%
\int_{0}^{\infty }{\frac{\xi ^{1/2}d\xi }{e^{-\beta (\mu-\varepsilon_0-\Delta)} \ e^{\xi }-1}}  \nonumber \\
&{=}&\frac{L^{3}(2m/\hbar ^{2})^{3/2}}{4\pi ^{2}\beta^{3/2}}%
g_{3/2}(z_{1})\Gamma (3/2)
\label{Ne2}
\end{eqnarray}
with $z_{1} \equiv \exp[\beta(\mu - \varepsilon_0 - \Delta ]\leq 1.$ %Note the minus in the argument of the exponential which contrasts with the fugacity definition $z \equiv \exp[\beta \mu]$. 

%\begin{center}
%\includegraphics[height=2.10in,angle=0]{RelDispersion2pdf}
%\end{center}

%FIGURE 1
\begin{figure}[tbh]
\centerline{\epsfig{file=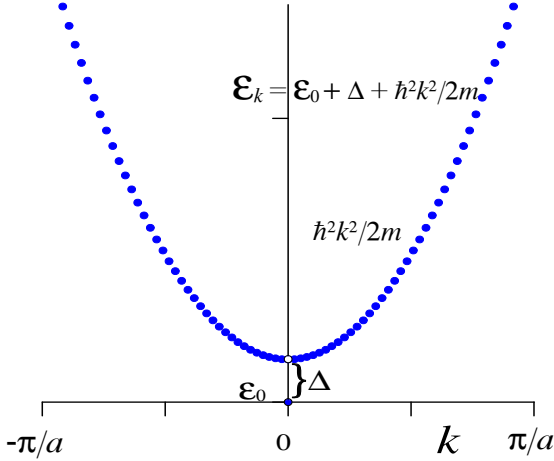,height=2.6in,width=3.0in}}
\caption{Particle energy.}
\label{fig:RelDis}
\end{figure}

\subsection{CBE critical temperature}

The critical temperature $T_c$ is the smaller temperature for which   
 $N_{e}(T_{c}) = N$ and 
as a consequence $\mu(T_c) =\varepsilon _{0}$, therefore
\begin{equation}
N = \frac{L^{3}(2m/\hbar^{2})^{3/2}}{4\pi ^{2}\beta _{c}^{3/2}}%
g_{3/2}(z_{1c})\Gamma (3/2)
\label{Tc}
\end{equation}
with $z_{1c} \equiv \exp[-\beta _{c}\Delta ]\leq 1.$ 
For $\Delta =0 $, $z_{1c}=1$ and we recover the critical temperature $T_{0}$ of a ideal Bose gas via the relation
\begin{equation}
N = \frac{L^{3}(2m/\hbar^{2})^{3/2}}{4\pi ^{2}\beta _{0}^{3/2}}%
\zeta (3/2) \Gamma (3/2)
\label{T0}
\end{equation} 
with $\beta _{0}=1/k_BT_0$ and $T_0$ the BEC critical temperature for a ideal Bose gas with a number density equal to that of our system. Note that in this case the dispersion relation for the Bose gas starts at $\varepsilon_0$ but has no effect on the critical temperature as this is a translation of the starting point of the particle energies. 

For $\Delta \neq 0$, from the ratio between (\ref{Tc}) and (\ref{T0}),  
\begin{eqnarray*}
1 &=&\frac{\beta _{0}^{3/2}g_{3/2}(z_{1c})}{\beta _{c}^{3/2}\zeta (3/2)}%
\qquad \\
{\mbox{or} \qquad }T_{c} &=&\left( \frac{\zeta (3/2)}{g_{3/2}(z_{1c})}\right)
^{2/3}T_{0}
\end{eqnarray*}
%
%FIGURE 2
\begin{figure}[tbh]
\centerline{\epsfig{file=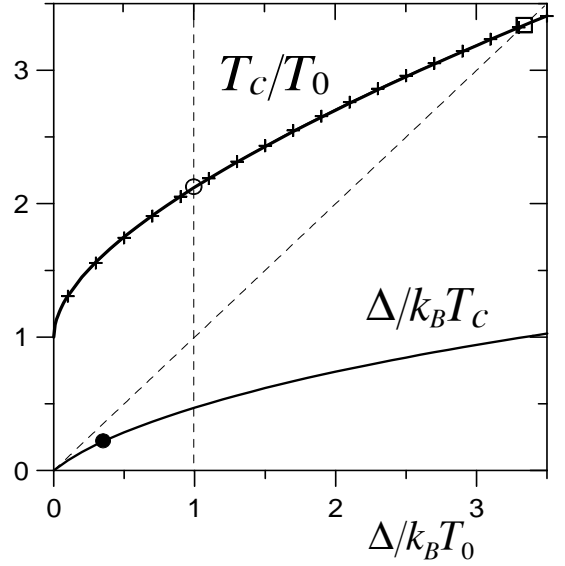,height=3.0in,width=3.0in}}
\caption{$Tc/T_0$ and $\Delta/k_BT_c$ as functions of the gap in units of $k_B T_0$. Crosses are the approximation (\ref{TcvsGapApprox}) for small $\Delta/k_BT_0$. Dot, circle and square mean $\Delta/k_B T_c =0.22$, $\Delta=k_B T_0$ and $\Delta = k_B T_c$, respectively.}
\label{fig:TcvsGap}
\end{figure}
Considering that $\zeta(3/2)/g_{3/2}(z) \simeq 1 + (3/2) \ (1 - z)^{1/2} + (9/2) \ (1 - z)^{3/2}$ for $z \in [0.8, 1]$ and $g_{3/2}(z) \simeq \exp[z] - 1$ for  $z \in [0.0, 0.8]$, and that for  
$\Delta > 0$, $T_c/T_0 > 1$ and $z_{1c}= \exp[-\Delta/k_B T_c] = \exp[-(\Delta/k_B T_0)T_0/T_c]$ then $g_{3/2}(z_{1c})$ can be approximated by 
\begin{equation}
%\hspace{-1.0cm} g_{3/2}(z_{1c}) \simeq
  \left\{
\begin{array}{l}
{\zeta(3/2)}/{(1 + (3/2) \ (1 - z_{1c})^{1/2} +  (9/2) (1 - z_{1c})^{3/2})}  \\ 
{\ \ \mbox{if}\ \ \ }\Delta/k_B T_0 \in [0, \ 0.22 \ T_c/T_0] \\ \\
\exp[z_{1c}] - 1 {\ \ \ \mbox{if} \ \ \ }\Delta/k_B T_0 \in [0.22 \ T_c/T_0, \ \infty]
\end{array}%
\right.  
\label{ec:aproxTc}
\end{equation}
Then, for small values of $\Delta/k_BT_0 < 0.22 \ T_c/T_0$ or $\Delta/k_BT_c < 0.22 $, the critical temperature takes the following form
\begin{equation}
T_{c} \simeq \left[ 1 + 3/2 \ (1 - z_{1c})^{1/2} +
 9/2 (1 - z_{1c})^{3/2}\right]^{2/3} T_{0}
\end{equation}
which it can be approximated after expansion around $\Delta/k_BT_c =0$ 
by
\begin{equation}
\frac{T_{c}}{T_0} \simeq 1 + (\Delta/k_BT_c)^{1/2} - (\Delta/k_BT_c)/4 + \frac{35}{12} (\Delta/k_BT_c)^{3/2} + ....
\end{equation}
We use the structure of the last expansion to propose 
\begin{equation}
\frac{T_{c}}{T_0} \simeq 1 + (\Delta/k_BT_c)^{1/2} - A(\Delta/k_BT_c) + B (\Delta/k_BT_c)^{3/2}
\label{TcvsGapApprox}
\end{equation}
with $A = 0.0277$ and $B = 1.3099$ such that the last expression reproduces the values at the open circle and square of Fig. \ref{fig:TcvsGap}. In this Fig.  \ref{fig:TcvsGap} crosses represent expression (\ref{TcvsGapApprox}). In the same figure we plot $\Delta/k_BT_c$ as a function of $\Delta/k_BT_0$, where the dot means $\Delta/k_BT_c = 0.22$.  

%..{\huge OJO!!! para BCS $2\Delta_0/k_BT_0=3.54$}. 

For values of $\Delta/k_BT_0 > 0.22 \ T_c/T_0$ or $\Delta/k_BT_c > 0.22 $ 
%(see big dot in Fig. \ref{fig:TcvsGap} {\huge OJO!!!})
 the critical temperature can be approximated using
%obtained from the following relation
%
%
%\vspace{0.2cm}
%NOTAR que $0.22 = \ln{0.8}$. Antes hab\'ia tomado $\ln{3/4}$.
%\vspace{0.2cm}
%
% O aproximando 
$g_{3/2}(z_{1c})\simeq \exp[\exp[-\beta
_{c}\Delta ]]-1$, then
\begin{equation}
T_{c} \simeq \left( \frac{\zeta (3/2)}{{\exp}[{\exp}[-\beta
_{c}\Delta ]]-1}\right)
^{2/3}T_{0} \\
\end{equation}
and
\begin{equation}
\Delta/k_BT_0 \simeq - (T_c/T_0) \ln(\ln[1+\zeta(3/2)/(T_c/T_0)^{3/2}]).
\end{equation}
Special values of the critical temperature are those for $\Delta/k_BT_c = T_0/T_c$ and $\Delta/k_BT_c = 1$. These values are $T_c/T_0 = 2.1228$ and 3.3376, which are shown in Fig. \ref{fig:TcvsGap} as  open circle and square, respectively.\\

\subsection{Condensed fraction}

For $T <  T_{c}$ the number of particles $N_0$ in the energy ground state is a ``large" fraction of the total number $N$ of particles in the system.   
%
%\begin{equation*}
%N=N_{0}+N_{e}=N_{0}+\sum_{\mathbf{k\neq 0}}{\frac{1}{e^{\beta _{c}\Delta
%}e^{\beta \lbrack \hbar ^{2}k^{2}/2m]}-1}},
%\end{equation*}
%
The condensed fraction is $N_0/N= (N-N_e)/N = 1- N_e/N$, where using the ratio between (\ref{Ne1}) and (\ref{Tc}) we arrived to  
\begin{equation}
N_{0}/N = 1-\beta _{c}^{3/2}g_{3/2}(z_{0})/[\beta ^{3/2}g_{3/2}(z_{0c})],
\end{equation}
or
\begin{equation}
N_{0}/N=1-\frac{T^{3/2}}{T_{c}^{3/2}}\frac{g_{3/2}(z_{0})}{g_{3/2}(z_{0c})} \xrightarrow[\Delta \to 0]{}  1- (T/T_0)^{3/2},
\label{fraccond}
\end{equation}%
with $z_0 \equiv Exp[-\beta \Delta] = z_1$ for $T < T_c$, and at $T = T_c$ $z_{0c}= z_{1c}$.  
%\pagebreak
%
%\bigskip QUITAR \ Then
%\begin{equation*}
%d(1-N_{0}/N)/dT=d\left( \frac{T^{3/2}}{T_{c}^{3/2}}\frac{g_{3/2}(z_{0})}{%
%g_{3/2}(z_{0c})}\right) /dT=\frac{(3/2)g_{3/2}(z_{0})T^{1/2}}{%
%T_{c}^{3/2}g_{3/2}(z_{0c})}+\frac{T^{3/2}g_{1/2}(z_{0})\Delta k_{B}\beta ^{2}%
%}{T_{c}^{3/2}g_{3/2}(z_{0c})},
%\end{equation*}
%

%\begin{figure}[]
%\includegraphics[height=2.20in,angle=0]{frac-condensada}
%\end{figure}

%FIGURE 3
\begin{figure}[tbh]
\centerline{\epsfig{file=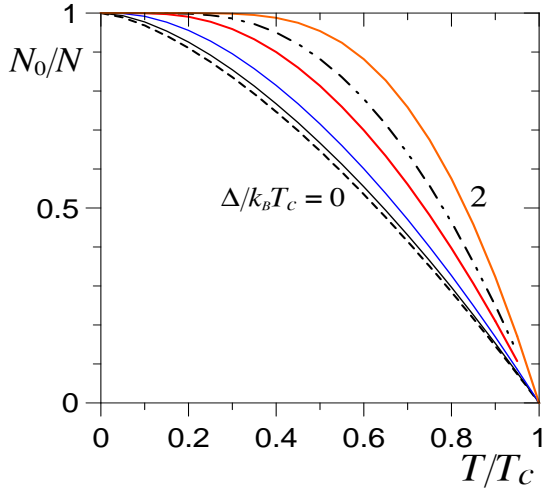,height=2.6in,width=3.0in}}
\caption{Condensed fraction $N/N_0$ for several gap values in units of $k_B T_c$, between 0 and 2.}
\label{fig:FracCon}
\end{figure}

When $\Delta = 0 $ we recover the condensed fraction for a IBG, whereas when the gap increases 
a higher temperature is required to diminish the condensed fraction from one. In other words, the particles in the energy ground state  require a larger energy to reach the next energy level. See Fig. \ref{fig:FracCon}.
 
\section{Thermodynamic properties}

\subsection{Internal energy}
The internal energy $U(L^{d},T)$ for bosons with energies  
 $\varepsilon _{k}$, is 
\begin{equation}
U(L^{d},T)=\sum_{\mathbf{k}}\varepsilon _{k}n_{k}\equiv \sum_{\mathbf{k}}%
\frac{\varepsilon _{k}}{e^{\beta (\varepsilon _{k}-\mu )}-1},
\end{equation}
Substituting (\ref{rdg}) in the last equation, in the thermodynamic limit we obtain 
\begin{equation*}
U(L^{d},T)=\mathbf{\varepsilon }_{0}N_{0}+\frac{L^{3}4\pi \,}{(2\pi )^{3}}%
\int_{0^{+}}^{\infty }\frac{(\mathbf{\varepsilon }_{0}+\Delta +\hbar
^{2}k^{2}/2m)k^{2}dk}{e^{\beta \lbrack \mathbf{\varepsilon }_{0}+\Delta
+\hbar ^{2}k^{2}/2m-\mu (T)]}-1},
\end{equation*}
Doing $\xi \equiv \beta \lbrack \hbar ^{2}k^{2}/2m]$,  $d$\bigskip $\xi
=(\beta \hbar ^{2}/2m)2kdk$ and $\xi ^{3/2}=(\beta \hbar ^{2}/2m)^{3/2}k^{3}$, last equation becomes
\begin{gather}
U(L^{d},T)=\mathbf{\varepsilon }_{0}N+\frac{L^{3}}{4\pi ^{2}}\;\frac{\Delta
}{\beta ^{3/2}}(2m/\hbar ^{2})^{3/2}\int_{0^{+}}^{\infty }\frac{\xi
^{1/2}d\xi }{z_{1}^{-1}e^{\xi }-1}  \notag \\
+\frac{L^{3}}{4\pi ^{2}}\frac{(2m/\hbar ^{2})^{3/2}}{\beta ^{5/2}}%
\int_{0^{+}}^{\infty }\frac{\xi ^{3/2}d\xi }{z_{1}^{-1}e^{\xi }-1},
\end{gather}
%with $z_{1}\equiv \exp[\beta (\mu -\mathbf{\varepsilon }_{0}-\Delta )].$
%
%
or
\begin{gather}
U(L^{d},T)/N=\mathbf{\varepsilon }_{0}+\Delta (1-N_{0}/N)+  \notag \\
\frac{L^{3}}{4\pi ^{2}}\frac{(2m/\hbar ^{2})^{3/2}}{\beta ^{5/2}N}\Gamma
(5/2)g_{5/2}(z_{1}),
\end{gather}

Dividing both sides of the last equation by $k_BT$ and using Eq. (\ref{Tc}) for $N$ in the last term of the second member, we have
%
%
%{\tiny{
%
%
%Como
%\[
%N=\frac{L^{3}(2m/\hbar ^{2})^{3/2}}{4\pi ^{2}\beta _{c}^{3/2}}%
%g_{3/2}(z_{0c})\Gamma (3/2)
%\]%
%
%
%Escribimos la energ\'{\i}a interna como
%
%\begin{equation}
%U(L^{d},T)/N=\mathbf{\varepsilon }_{0}+\Delta (1-N_{0}/N)+\frac{\beta
%_{c}^{3/2}}{\beta ^{5/2}}\frac{\Gamma (5/2)g_{5/2}(z_{1})}{\Gamma
%(3/2)g_{3/2}(z_{0c})},
%\end{equation}
%
%\bigskip or%

\begin{equation}
\frac{U(L^{d},T)}{N k_B T}= \frac{\varepsilon_{0}}{k_BT} + \frac{\Delta (1-N_{0}/N)}{k_BT} + \frac{\Gamma
(\frac{5}{2}) g_{5/2}(z_{1})}{\Gamma (\frac{3}{2})g_{3/2}(z_{1c})} \left(\frac{T}{T_c} \right)^{3/2},
\end{equation}
where we can see that the internal energy has a first term which is the a reference energy.
If we use the relation between the grand potential $\Omega$ and $PV$we are able to arrive to the equation of state 
\[  PV = (2/3) U - (2/3)N [\varepsilon_0 + (1- N_0/N) \Delta)]
\]
which reduces to the well known case when $\Delta$ and $\varepsilon_0$ are zero. 
 
%
%COMENTAR ALGO SOBRE LA ENERGIA INTERNA???

\subsection{Specific heat}
The specific heat expression is obtained from the temperature derivative of the internal energy %
\begin{gather}
C_{V}/N=d(U/N)/dT=\Delta d(1-N_{0}/N)/dT+  \notag \\
\frac{\Gamma (5/2)\beta _{c}^{3/2}}{\Gamma (3/2)g_{3/2}(z_{0c})\beta ^{3/2}}%
(5/2)k_{B}g_{5/2}(z_{1})  \notag \\
+\frac{\Gamma (5/2)\beta _{c}^{3/2}}{\Gamma (3/2)g_{3/2}(z_{0c})\beta ^{5/2}}%
g_{3/2}(z_{1})\times   \notag \\
[-k_{B}\beta ^{2}(\mu -\mathbf{\varepsilon }_{0}-\Delta )+\beta d\mu /dT],
\end{gather}
where we have used 
\begin{equation*}
d/dT(\beta \lbrack \mu -\mathbf{\varepsilon }_{0}-\Delta ])=-k_{B}\beta
^{2}[\mu -\mathbf{\varepsilon }_{0}-\Delta ]+\beta d\mu /dT.
\end{equation*}
and that the temperature derivative of $\mu$,  is 
\begin{eqnarray*}
&&(3/2)k_{B}\beta ^{-1/2}g_{3/2}(z_{1})+\beta ^{-3/2}g_{1/2}(z_{1}) \\
&&\times \lbrack -k_{B}\beta ^{2}(\mu -\mathbf{\varepsilon }_{0}-\Delta
)+\beta (d\mu /dT)].
\end{eqnarray*}

For $T\leq T_{c},$ $\mu =\mathbf{\varepsilon }_{0}$ and $d\mu /dT=0$%
, then
\begin{gather}
C_{V}^{-}/Nk_{B}=\frac{(3/2) g_{3/2}(z_{0})(T/T_c)^{1/2}(\Delta/k_BT_c)}{g_{3/2}(z_{0c})}+  \notag \\
\frac{g_{1/2}(z_{0})(\Delta/k_B T_c)^{2}}{g_{3/2}(z_{0c})(T/T_{c})^{1/2}}%
+\frac{(5/2)\Gamma (5/2)g_{5/2}(z_{0}) (T/T_{c})^{3/2}}{\Gamma
(3/2)g_{3/2}(z_{0c})}  \notag \\
+\frac{\Gamma (5/2)g_{3/2}(z_{0})(T/T_c)^{1/2}(\Delta/k_BT_c)}{\Gamma (3/2)g_{3/2}(z_{0c})}
\end{gather}

For $T\geq T_{c},$ $N_{0}\simeq 0$ and
\begin{gather*}
C_{V}^{+}/Nk_{B}=\frac{(5/2)\Gamma (5/2)g_{5/2}(z_{1})(T/T_c)^{3/2}}{%
\Gamma (3/2)g_{3/2}(z_{0c})} \\
-\frac{(3/2)\Gamma (5/2)g_{3/2}^{2}(z_{1})(T/T_c)^{3/2}}{\Gamma
(3/2)g_{3/2}(z_{0c})g_{1/2}(z_{1})}
\end{gather*}
or
\begin{equation}
C_{V}^{+}/Nk_{B}=\frac{3(T/T_c)^{3/2}}{
2 \ g_{3/2}(z_{0c})} \left[\frac{5}{2}g_{5/2}(z_{1})-  
\frac{3 \ g_{3/2}^{2}(z_{1})}{2 \ g_{1/2}(z_{1})} \right]
\end{equation}

Note that for all temperature regimes, the specific heat is $\mathbf{\varepsilon }_{0}$ independent. However it is $\Delta$ dependent for temperatures below $T_c$ only. For $T \to 0$ and $\Delta \neq 0$, $z_0 \to 0$ and we can approximate $g_{\sigma}(z_0) \simeq z_0$ by $\sigma = 5/2, 3/2 \, \mbox{or} \, 1/2$. Then   
\begin{gather}
C_{V}^{-}/Nk_{B}\simeq \exp(-\Delta/k_BT) \left\{ \frac{(3/2) (T/T_c)^{1/2}(\Delta/k_BT_c)}{g_{3/2}(z_{0c})}+ \right. \notag \\
\frac{(\Delta/k_B T_c)^{2}}{g_{3/2}(z_{0c})(T/T_{c})^{1/2}}%
+\frac{(5/2)\Gamma (5/2) (T/T_{c})^{3/2}}{\Gamma
(3/2)g_{3/2}(z_{0c})}  \notag \\
\left. +\frac{\Gamma (5/2)(T/T_c)^{1/2}(\Delta/k_BT_c)}{\Gamma (3/2)g_{3/2}(z_{0c})} \right\},
\end{gather}
or
\begin{gather}
C_{V}^{-}/Nk_{B}\simeq \frac{\exp(-\Delta/k_BT)(T/T_c)^{3/2}}{g_{3/2}(z_{0c})} 
\left\{\frac{15}{4} + \right. \notag \\
\left.  3 (\Delta/k_BT) + (\Delta/k_B T)^{2}  \right\}
\end{gather}
i.e., for very low temperatures the temperature dependence of the specific heat is exponential. 

However, when $\Delta \to 0$ we recover the well known behavior of $C_V$ for an ideal Bose gas at low temperatures, i.e. it is proportional to $T^{2/3}$.
%AQUI REORDENO LA ECUACION ANTERIOR
%\begin{gather}
%C_{V}^{-}/Nk_{B}\simeq \frac{\exp(-\Delta/k_BT)(T/T_c)^{3/2}}{g_{3/2}(z_{0c})} 
%\left\{\frac{15}{4} + \right. \notag \\
%\left. \frac{ 3 (\Delta/k_BT_c)}{(T/T_c)}+ 
% \frac{(\Delta/k_B T_c)^{2}}{(T/T_{c})^{2}}  \right\}
%\end{gather}
%
%\begin{center}
%%\begin{figure}[htb]
%\includegraphics[height=2.20in,angle=0]{CvvsTGap}
%%\end{figure}
%\end{center}
%

%FIGURE 4
\begin{figure}[tbh]
\centerline{\epsfig{file=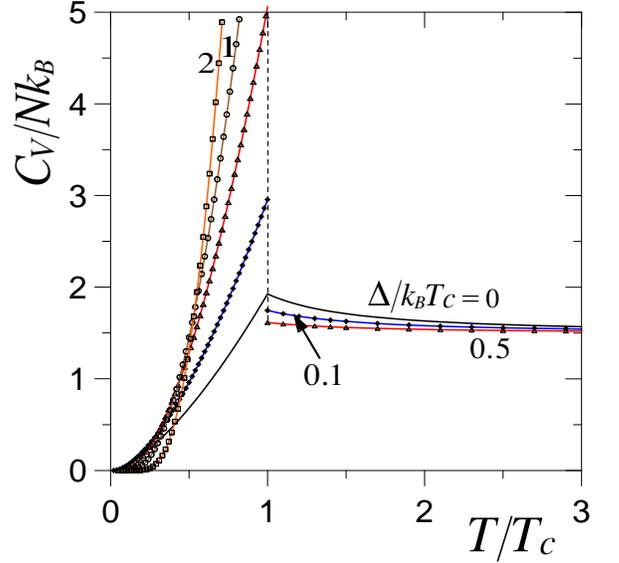,height=3.0in,width=3.2in}}
\caption{Specific heat as a function of $T/T_c$ for $\Delta/k_B T_c =$ 0, 0.1, 0.5, 1 and 2. For $\Delta/k_B T_c =$ 1 and 2, the $T/T_c > 1$ branches are not shown.}
\label{fig:SpecHeat}
\end{figure}

\vspace{1.0cm}

%FIGURE 5
\begin{figure}[tbh]
\centerline{\epsfig{file=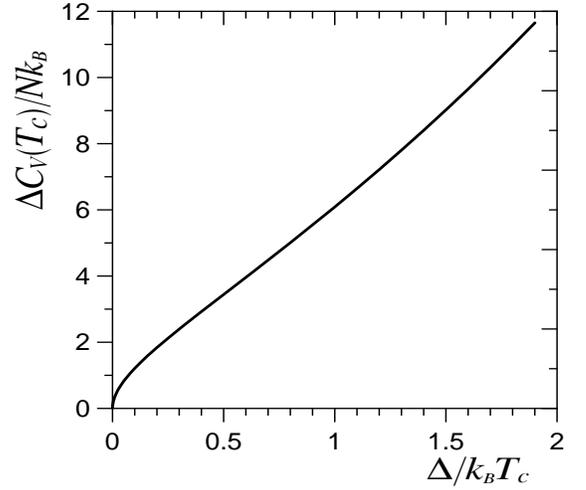,height=2.6in,width=3.0in}}
\caption{Specific heat jump.}
\label{fig:SaltoSpecHeat}
\end{figure}

\subsection{Jump magnitude in the specific heat}
For $T=T_{c},$ $\mu =\mathbf{\varepsilon }_{0}$ y \ $z_{1}\rightarrow
z_{0c} $
\begin{gather}
(C_{V}^{-}-C_{V}^{+})/Nk_{B}|_{T_{c}}= 3 (\Delta /k_{B}T_{c}) \nonumber \\
%+\frac{(5/2)\Gamma (5/2)\beta _{c}^{3/2}g_{5/2}(z_{0c})}{\Gamma
%(3/2)g_{3/2}(z_{0c})\beta ^{3/2}}
%+\frac{\Gamma (5/2)\beta _{c}^{3/2}\beta
%^{2}\Delta }{\Gamma (3/2)\beta ^{5/2}} 
%-\frac{\Gamma (5/2)\beta _{c}^{3/2}(5/2)g_{5/2}(z_{0c})}{\Gamma
%(3/2)g_{3/2}(z_{0c})\beta ^{3/2}}
+\frac{g_{1/2}(z_{0c})(\Delta /k_{B}T_{c})^{2}}{g_{3/2}(z_{0c})}
+\frac{9}{4} \frac{g_{3/2}(z_{0c})}{g_{1/2}(z_{0c})} \nonumber
\end{gather}
%
%\begin{center}
%%\begin{figure}[htb]
%\includegraphics[height=2.0in,angle=0]{SaltoCvvsDel}
%%\end{figure}
%\end{center}

In Fig. \ref{fig:SaltoSpecHeat} we can see the specific heat jump which it is an increasing function of  $Delta/k_BT_c$. 
For small $\Delta$ the specific heat jump can be approximate by 
$(C_{V}^{-}-C_{V}^{+})/Nk_{B}|_{T_{c}} \simeq (10/3)(\Delta/k_BT_c)^{1/2}$.

\section{Conclusions}
 After calculating the BEC critical temperature, the condensed fraction, the internal energy, the isochoric specific heat and its jump magnitude at $T_c$, for an interactionless Bose gas whose particle energies have a gap between the ground energy and the first excited energy level,
%  staare given by Ec. (\ref{rdg}), 
we conclude the following:
\begin{itemize}
\item All the calculated thermodynamic properties are independent of the magnitude of the ground state energy $\varepsilon_0$,
%\item %-independent, 
except the internal energy which is measured from the reference energy of magnitude $N \varepsilon_0$, instead of the zero ground state energy of a ideal Bose gas. 
%$\varepsilon_0$ if its momentum is zero, and otherwise    proportional to their squared momentum when their momentum is larger than zero while   for my  
%\item The magnitud of $\varepsilon_0$ no afecta el valor de la $T_c$ ni del calor espec\'ifico, s\'olo incrementa la energ\'ia interna por $N \varepsilon_0$ para toda $T$.
\item The gap notoriously increases the magnitude of the critical temperature and brings on the appearance of a jump in the specific heat.
%  El gap incrementa la magnitud de la temperatura cr\'itica y genera un salto en el calor espec\'ifico.
\item Below the critical temperature and near to zero absolute,
the temperature dependence of the specific heat is exponential with an exponent proportional to the gap, however when the gap becomes zero the temperature dependence of the specific heat becomes that of the IBG.  
% For low temperaturesPara bajas temperaturas,  el gap ocasiona que el calor espec\'ifico se comporte exponencialmente con la temperatura en lugar de una ley de potencias $T^{3/2}$. 

\end{itemize}

{\bf Acknowlegment}. We thank  partial support from projects CONACyT  221030, and PAPIIT-UNAM IN111613.

\end{document}